\documentclass[11pt,a4paper,onecolumn]{article}
\usepackage{amsmath,amsfonts,amssymb,bm,mathrsfs}
\usepackage{graphicx,color,xcolor}
\usepackage[normalem]{ulem} 
\usepackage[utf8]{inputenc}
\usepackage[T1]{fontenc}
\usepackage{url}

\graphicspath{{./Figs/}}
\definecolor{MidRed}{rgb}{0.78,0,0}
\definecolor{MidGreen}{rgb}{0,0.50,0}
\definecolor{MidBlue}{rgb}{0,0,0.68}

\newcommand{\unit}[1]{\ensuremath{\mathrm{#1}}}

\newcommand{\Celsius}{\ensuremath{\mathrm{^\circ C}}}
\newcommand{\req}[1]{(\ref{#1})}

\newcommand{\Today}{March 25, 2020}
\def\pageheads{E. Rubiola et al., Artifacts and Errors\ldots\hfill~\Today\hfill}
\title{Artifacts and Errors in\\[0.3ex]Cross-Spectrum Phase Noise Measurements}
\author{Yannick Gruson$^{\exists}$, 
Adrian Rus$^\star$, Ulrich L. Rohde$^\Diamond$,\\[0.5ex] 
Alexander Roth$^\Box$, and Enrico Rubiola$^{\exists\,\nabla\,\Re}$\\[0.8em]
\small\sffamily
$\exists$ FEMTO-ST Institute, UBFC and CNRS, Besançon, France.\\[0.ex] 
\small\sffamily
$\star$ Broadhurst, Bucharest, Romania.  Call sign YO3HHZ.\\[0.ex]
\small\sffamily
$\Diamond$ Synergy Microwave Corp., Paterson, NJ, USA\@.  Call sign N1UL.\\[0.ex]
\small\sffamily
$\Box$ Rohde \& Schwarz, München, Germany.\\[0.ex]
\small\sffamily
$\nabla$ Istituto Nazionale di Ricerca Metrologica INRiM, Torino, Italy\\[0.ex]
\small\sffamily
$\Re$ Reference author\\[0.8em]
\small\sffamily
\normalsize\sffamily\bfseries
http://rubiola.org}

\date{\vspace*{-0.3em}\small\sffamily\Today}

\begin{document}

\maketitle

\begin{abstract}
This article deals with the erratic and inconsistent phase-noise spectra often seen in low-noise oscillators, whose floor is of the order of $-180$ dBc/Hz or less.  Such oscillators are generally measured with two-channel instruments based on averaging two simultaneous and statistically independent measures.
Our new method consists of inserting a dissipative attenuator between the oscillator under test and the phase-noise analyzer. 
The thermal noise of the attenuator introduces a controlled amount of phase noise.  We compare the phase noise floor to the theoretical expectation with different values of the attenuation in small steps.
The analysis reveals a negative bias (underestimation of phase noise) due to the thermal energy of the internal power splitter at the instrument input, and an uncertainty due to crosstalk between the two channels.  In not-so-rare unfortunate cases, the bias results in a negative phase-noise spectrum, which is an obvious nonsense.
Similar results are observed separately in three labs with instruments from the two major brands.  We give experimental evidence, full theory, and suggestions to mitigate the problem.
Our multiple-attenuators method provides quantitative information about the correlation phenomena inside the instrument.
\end{abstract}

\section{Introduction}
Modern  analyzers measure the phase noise (PN, or PM noise) by correlation and averaging on the simultaneous measurement of the oscillator under test (DUT) with two separate channels, each consisting on a phase detector and a frequency reference.  The DUT noise is extracted after rejecting the background noise of the instrument, assuming that the two channels are statistically independent.  After the seminal paper \cite{Walls-1976}, and the early application shown in \cite{WWalls-1992}, this choice is adopted by virtually all manufacturers (Table~\ref{tab:PN-Analyzers}).
The dual channel scheme comes in two flavors, with one or two reference oscillators. We focus on the latter because it enables the noise rejection of the reference oscillators, and also of the frequency synthesizers which may be interposed between reference oscillators and phase detectors.

The correlation-and-averaging process rejects the single-channel noise proportionally to $1/\sqrt{m}$, where $m$  is the number of averaged spectra, that is, 5 dB per factor-of-ten.  Nowadays, digital electronics provides a high computing power and memory size for cheap, as compared to the cost and to the complexity of RF and microwave technology.  Thus, the theoretical rejection can exceed 30 dB if the experimentalist accepts the long measurement time it takes, ultimately limited by the time-frequency indetermination theorem.  However, such rejection cannot be achieved in practice because of fundamental phenomena and artifacts.  The thermal energy in the input power splitter \cite{Rubiola-2000-RSI,Ivanov-2002,Nelson-2014,Hati-2016,Hati-2017}
and impedance matching \cite{Breitbarth-2013} first caught the attention of the scientific community.  
These and other problems were addressed in three international workshops \cite{L(f)-2017,L(f)-2015,L(f)-2014}.

Most practitioners, naively, believe that a noise analyzer always \emph{over-estimates} the DUT noise because it \emph{adds} its own background noise. This is not true in the case of the two-channel instruments because the cross spectrum is the frequency-domain equivalent of the covariance. The correlation between channels introduces systematic errors and artifacts, which can be positive or negative. The consequence is that there is no a-priori rule to state whether the instrument over-estimates or under-estimates the DUT noise.  A problem is that the noise rejection due to averaging, usually calculated and displayed together with the phase noise, does not account for artifacts and systematic errors.  Another problem is that the instruments display the absolute value of the cross spectrum, giving no warning about negative outcomes.  The combination of these facts originates erratic and misleading results.

\begin{figure}[t]
\centering
\includegraphics[width=3.5in]{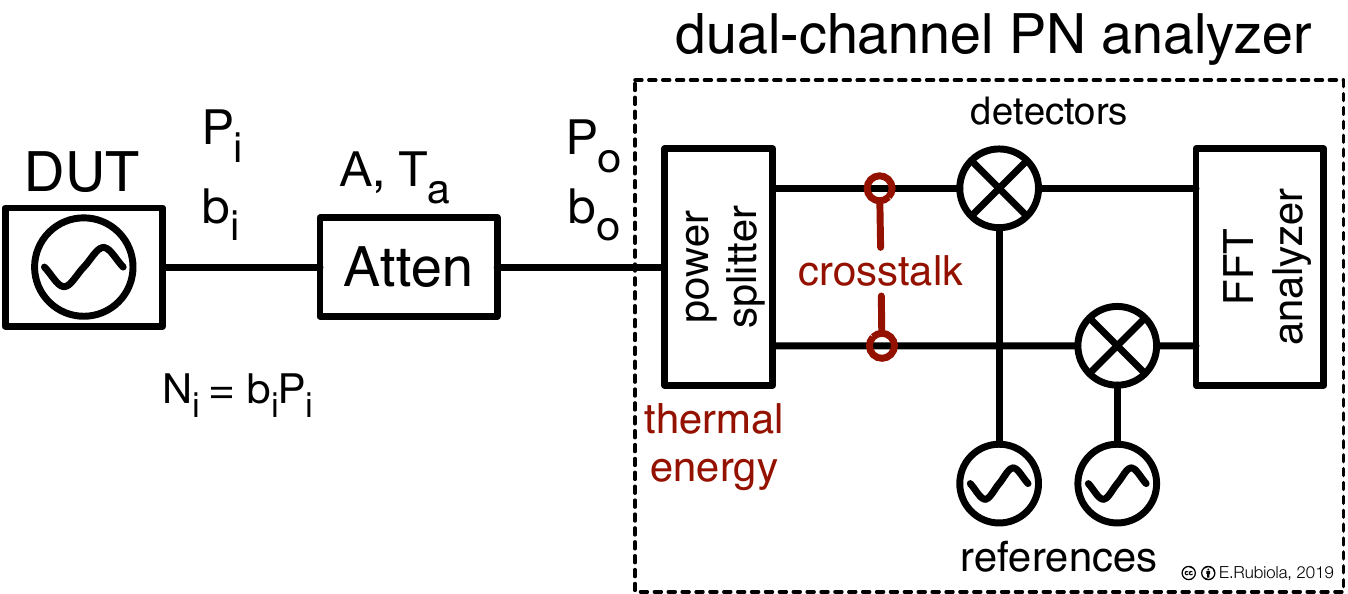}
\caption{Block diagram of the experiment.}
\label{fig:Experiment}
\end{figure}

We propose an experiment (Fig.~\ref{fig:Experiment}) that reveals the presence systematic errors due to unwanted correlated terms.  We focus on the 100 MHz OCXOs because this type of oscillator exhibits the lowest white PM noise floor.  However trivial the experiment may seem, nothing even broadly similar has been reported in the literature.  We provide all the details related to two specific cases, together with the full theoretical interpretation. 

\begin{table}[t]\centering
\caption{Dual-Channel Phase Noise Analyzers}
\label{tab:PN-Analyzers}
\sffamily\small
\begin{tabular}{ll}\hline
\bfseries Brand & \bfseries Type or Series\\\hline
AnaPico& APPH series\\
Berkeley Nucleonics Corp\@. & 7000 series\\
Holzworth  & HA7062 series\\
Jackson Labs Technologies & PhaseStation 53100A\\
Keysight Technologies & E5500 series\\
Microsemi Corporation & 3120A / 5120A \\
NoiseXT & DCNTS / NXA \\
OEwaves & HI-Q TMS\\
Rohde \& Schwarz & FSWP series 
\end{tabular}
\end{table}

\section{Phase Noise and Thermal Energy}\label{sec:PM-Noise}%
Let us start with a review of key facts, based on References \cite{Rubiola-2000-RSI,Ivanov-2002,Nelson-2014,Hati-2016,Hati-2017}.
The phase noise is described in terms the power spectral density (PSD) of the random phase $\varphi(t)$, and denoted with $S_\varphi(f)$.  
A model that is found useful to describe oscillators and components is the polynomial law
\begin{equation}
\label{eqn:plaw}
S_\varphi(f) = \sum_{n\leq-4}^{0}
\mathsf{b}_{n}f^{n} 
~~~\unit{[rad^2/Hz]}~,
\end{equation}
where the term $\mathsf{b}_{0}$ is the white PM noise, $\mathsf{b}_{-1}/f$  is the flicker PM noise, $\mathsf{b}_{-2}/f^{2}$ is the white FM noise, $\mathsf{b}_{-3}/f^{3}$ is the flicker FM noise, $\mathsf{b}_{-4}/f^{4}$ is the frequency random walk, and other terms can be added.  
The quantity $\mathscr{L}(f)$, most often used by the manufacturers, is  defined as $\mathscr{L}(f) = (1/2) S_\varphi(f)$ \cite{IEEE-STD-1139-2008}.  
As a matter of fact, white phase noise is mostly of additive origin.  Accordingly, it can be written as
\begin{equation}
\label{eqn:b0}
\mathsf{b}_{0}=\frac{N}{P}~,
\end{equation}
where  $P$  is the carrier power, and  $N$  is the power spectral density (PSD) of the RF noise.  In this context, we prefer the unit W/Hz to  J\@.  
By analogy with the PSD $N=kT$ of the thermal noise,
we associate to $\mathsf{b}_{0}$ the equivalent temperature
\begin{equation}
\label{eqn:Teq}
T=\frac{P\,\mathsf{b}_{0}}{k},
\end{equation}
where $k=1.380649{\times}10^{-23}$ J/K (exact) is the Boltzmann constant.

\subsection{The Effect of the Attenuator}
Physical insight suggests that the dissipative attenuator can only degrade the signal-to-noise ratio (SNR), which results in increased PM noise.
Focusing on the white noise at the attenuator in and out, we use the subscripts ``$i$''  and ``$o$'' dropping the subscript 0.  For example, $\mathsf{b}_{i}$ stands for $\mathsf{b}_{0\,i}$, and $\mathsf{b}_{o}$ for $\mathsf{b}_{0\,o}$.  Thus \req{eqn:b0} rewrites as $\mathsf{b}_{i}=N_{i}/P_{i}$, or $\mathsf{b}_{o}=N_{o}/P_{o}$.
Assuming that everything is matched to the characteristic impedance $R_0$, the RF white noise at the attenuator output is
\begin{equation}
\label{eqn:Atten}
N_{o} = kT_iA^2 + kT_a\left(1-A^2\right)~,
\end{equation}
where  $kT_{i}$ is the input noise, $A$ is the voltage gain of the attenuator, $A^2<1$, and  $T_a$ is the temperature of the attenuator.  The term  $kT_iA^2$  means that the input noise  $kT_i$ is attenuated by the factor $A^2$, like any signal.  The term  $kT_a(1-A^2)$  is the thermal noise added by the attenuator.  This is obvious if one replaces the oscillator with a resistive load $R_0$  at the temperature $T_a$.  In this condition the output is equivalent to a resistor $R_0$ at the temperature $T_a$.  Thus, the total output noise is $kT_a$, independent of  $A$. 
Equation \req{eqn:Atten} is well known in radio astronomy, where it finds application in the estimation of the effect of losses in the antenna and in the line between antenna and receiver \cite[Sec.~7-2b (Noise Temperature of an Attenuator)]{Tiuri-1966}, and in the calibration of the receiver \cite[Sec.~4.2.4 (Receiver Calibration)]{Rohlfs-2000}.

After \req{eqn:b0}--\req{eqn:Atten}, the white PM noise at the attenuator output is
\begin{equation}
\label{eqn:b0-Atten}
\mathsf{b}_o=\frac{kT_i}{P_i} + \frac{kT_a\left(1-A^2\right)}{A^2P_i},
\end{equation}
which is obviously greater than $\mathsf{b}_i=kT_i/P_i$.

\section{Inside the Dual-Channel Noise Analyzer}

\subsection{The Cross-Spectrum Estimator}\label{sec:Cross-spectrum}

The cross-spectrum estimator is a general tool.  Here, it finds application to two key signals inside the instrument: (i) the voltage at the outputs of the power splitter, and (ii) the phase at the output of the detectors (Fig:~\ref{fig:Experiment}).  Let us start with   
\begin{align}
x(t) &= a(t) + c(t) \label{eqn:x}\\
y(t) &= b(t) + c(t) \label{eqn:y}
\end{align}
where $a(t)$ and $b(t)$ are the background noise of the channel $\mathcal{A}$ and $\mathcal{B}$.  They are statistically independent, and have zero mean and equal or similar variance.  The signal $c(t)$ is the target, that is, the DUT noise.  Thus, the statistical properties of $c(t)$ are measured after averaging out $a(t)$ and $b(t)$.
As said, \req{eqn:x}-\req{eqn:y} apply to the RF signal or to the PM noise, at choice.  Thus, we can measure the RF spectrum [W/Hz], or the PM noise spectrum [\unit{rad^2/Hz}]. 
The reader interested to know more about the method should refer to \cite{Rubiola-2010-arXiv}, \cite{Sampietro-1999} and \cite{WWalls-1992}.

As a mathematical concept, the cross PSD is defined as the Fourier transform of the autocorrelation function.  However, under certain conditions which are generally met in the case of physical signals digitized on a finite acquisition time $T$, it can be evaluated using the Fourier transforms. 
Thus, denoting the discrete Fourier transform with the uppercase letter, as in $X(f) \leftrightarrow x(t)$, the one-sided cross PSD is evaluated as
\begin{equation}
S_{yx}(f) =\frac{2}{T} \Bigl[Y(f)\,X^{\ast}(f)\Bigr]~.
\label{eqn:Syx}
\end{equation}
The denominator $T$ is the acquisition time, the superscript ``$\ast$''  means complex conjugate, and the factor ``2'' accounts for energy conservation after suppressing the negative frequencies.
Equation \req{eqn:Syx} states a general fact, thus it holds for one realization, for the average or for the expectation, depending on the context.
Dropping the frequency and expanding $X=A+C$ and $Y=B+C$ we get
\begin{equation}
S_{yx}=\frac{2}{T}\Bigl(B+C\Bigr)\:\Bigl(A^{\ast}+C^{\ast}\Bigr)~.
\label{eqn:SyxSphi}
\end{equation}
The mathematical expectation $\mathbb{E}\left\{S_{yx}\right\}$ is
\begin{equation}
\mathbb{E}\bigl\{S_{yx}\bigr\} = 
\frac{2}{T}\mathbb{E}\bigl\{CC^\ast\bigr\} =
\mathbb{E}\bigl\{S_{c}\bigr\} \label{eqn:SyxE}
\end{equation}
because $\mathbb{E}\{BA^\ast\}=0$, $\mathbb{E}\{BC^\ast\}=0$, and $\mathbb{E}\{CA^\ast\}=0$.
All the useful information is in $CC^{\ast}$, thus $S_{c}>0$.  By contrast, all the background noise goes in $BA^{\ast}$, $BC^{\ast}$ and $CA^{\ast}$, and under normal circumstances it is equally distributed between real and imaginary part. It is therefore clear that the optimum estimator is
\begin{align}
\widehat{S_{yx}} &= \Re\left\{\left< S_{yx} \right>_{m} \right\}~.
\label{eqn:Re-estim}
\end{align}
This estimator has two important properties, (i) it is unbiased, and (ii) it is the fastest because it takes in the smallest amount of background noise.  A problem with $\Re\left\{\left<S_{yx}\right>_{m}\right\}$ is that it is not always positive before averaging out the background noise, or in the presence of spurs.  The negative outcomes cannot be plotted on a logarithmic scale (dB).  The FSWP \cite[Equation (4)]{Feldhaus-2016} uses the estimator
\begin{align}
\widehat{S_{yx}} &= \left|\left< S_{yx}\right>_{m}\right|~.
\label{eqn:Abs-estim}
\end{align}
Albeit the documentation provided by other manufacturers gives little indication about the estimator, we believe that \req{eqn:Abs-estim} is the most chosen option.
A reason is that it shows no negative values, thus it is always suitable to be represented on a log scale.  Another reason is that such estimator is positively biased, and the bias decreases monotonically as $m$ increases.  Thus, under normal circumstances $\left|\right< S_{yx}\left>_{m}\right|$ converges to $\mathbb{E}\left\{S_{c}\right\}$, after decreasing monotonically during the measurement process.  The estimator \req{eqn:Abs-estim} matches the  behavior we observe in the regular use of the instruments, where no attenuator is inserted at the input.

Now we break the hypothesis of statistically independent channels, and we introduce the disturbing signal $d(t) \leftrightarrow D(f)$, the same in the two channels but for the sign, $\varsigma_{x}=\pm1$ and $\varsigma_{y}=\pm1$, 
as we did in \cite{Gruson-2017-TUFFC}
\begin{align}
X &= A + C + \varsigma_{x}D\label{eqn:XD}\\
Y &= B + C + \varsigma_{y}D\label{eqn:YD}~,
\end{align}
The signal $D$ is either correlated or anticorrelated.  Introducing $\varsigma=\varsigma_{x}\varsigma_{y}=\pm1$, and expanding $\mathbb{E}\left\{S_{yx}\right\}$ as above, we find
\begin{equation}
\mathbb{E}\bigl\{S_{yx}\bigr\} = 
\mathbb{E}\bigl\{S_c\bigr\} + 
\varsigma\mathbb{E}\bigl\{S_{d}\bigr\}~.
\label{eqn:SyxD}
\end{equation}
Thus, $\varsigma$ is the sign of the correlation coefficient, and the term $\varsigma\mathbb{E}\bigl\{S_{d}\bigr\}$ is a systematic bias, positive or negative.
A substantially equivalent approach is found in \cite[Section~II]{Nelson-2014}, which differs in the analysis of four separate cases (the presence or not of the disturbing signal $d(t)$, and the sign of the correlation), instead of the compact form \req{eqn:XD}--\req{eqn:YD} and~\req{eqn:SyxD}.

The disturbing signal can be (i) the thermal energy in the input power splitter (Sec.~\ref{sec:Power-splitter}), (ii) the crosstalk between the two channels (Sec.~\ref{sec:Crosstalk}), and (iii) the AM noise pickup \cite{Rubiola-2007-AM}, or other effects not considered here.

\subsection{Application to the Input Power Splitter}\label{sec:Power-splitter}
\begin{figure}[t]
\centering
\includegraphics[width=3.5in]{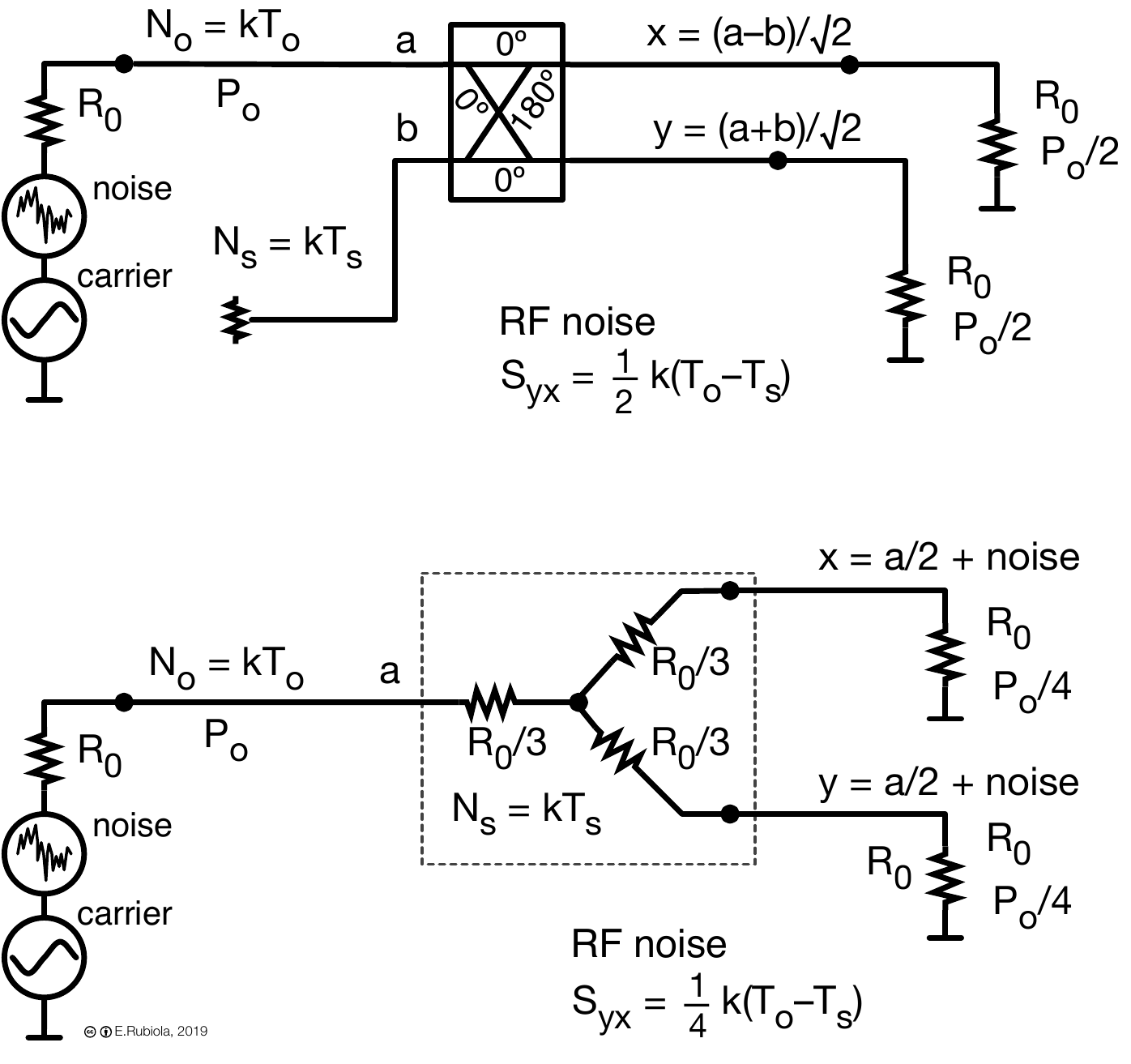}
\caption{Signal and noise model of the most common power splitters.  The reactive power splitter (top) is free from dissipation, thus it has 3 insertion loss.  The resistive power splitter (bottom) has 6 dB insertion loss.}
\label{fig:Power-splitters}
\end{figure}

Two types of power splitters are mostly used, shown on Fig~\ref{fig:Power-splitters}.  The loss-free splitter is a 3~dB directional coupler terminated at one input (dark port).  The resistive splitter is a Y network which attenuates the input signal by 6 dB\@.  Here, $x(t)$ and $y(t)$ are the RF voltages at the output of the power splitter.
Denoting with $T_o$ equivalent noise temperature at the power-splitter input, and with  $T_s$ the  temperature of the splitter, trite calculation 
shows that the correlated noise is
\begin{equation}
\mathbb{E}\left\{S_{yx}\right\}=\frac{1}{2}k\Bigl( T_{o}-T_{s} \Bigr) 
\label{eqn:Syx-Coupler}
\end{equation}
for the 3-dB dissipation-free coupler.  
Interestingly, \req{eqn:Syx-Coupler} is a classical result from Johnson thermometry \cite{Fischer-2015,White-1996}, with well known application in microwaves \cite{Allred-1962,Halford-1966}.

Similarly, we find
\begin{equation}
\mathbb{E}\left\{S_{yx}\right\}=\frac{1}{4}k\Bigl( T_{o}-T_{s} \Bigr) 
\label{eqn:Syx-Y}
\end{equation}
for the 6-dB resistive coupler. 
Deriving \req{eqn:Syx-Coupler} and \req{eqn:Syx-Y} from \req{eqn:SyxD}, $\varsigma$ does not need to appear explicitly because it always hold that $\varsigma=-1$.

Because the output power is $P_{o}/2$ for the 3-dB splitter and $P_{o}/4$ for the 6-dB splitter, the output SNR is the same, and the white PM noise is
\begin{equation}
\mathbb{E}\left\{\mathsf{b}_o\right\}=\frac{k(T_{o}-T_{s})}{P_o}~. 
\label{eqn:bo-Splitter}
\end{equation}
Reference \cite[Section IV]{Hati-2016} provides an extension to other less common types of power splitter.

\subsection{Hardware Architectures}\label{sec:Architectures}
The FSWP  \cite{PN:FSWP,Feldhaus-2016} is based on the SDR (Software Defined Radio) technology after down-converting the input to an appropriate IF (see \cite{Rohde-2017} for a modern treatise of SDR)\@.   
The mixers are used in the linear region because linearity prevents the AM noise from polluting the phase noise measurement.  The use of I-Q mixers enables to unwrap the phase, and to measure beyond the IF\@.  Two operating modes are used, depending on the Fourier frequency.  Up to 1 MHz, the input RF signal is down converted to 1.3 MHz.  Beyond 1 MHz, the reference synthesizers are set close to the input frequency, keeping the beat note below 10 Hz.  In both cases, $I$ and $Q$ of the down-converted signal are digitized, and the phase information is extracted in FPGA\@.  The FSWP uses a 3-dB coupler as the input power splitter (actually, three different couplers are switched, for $<1$   GHz, $1{-}8$ GHz, and $8{-}50$ GHz).

The E5052B \cite{PN:Keysight-E5500} is based on direct phase detection with double-balanced mixers as the phase-to-voltage converters.  The mixers are saturated at both inputs, and driven with synchronous signals kept in quadrature.  The mixer output is digitized and processed.  
The power splitter is a Y resistive network.  

In both cases, the signals $x(t)$ and $y(t)$ used to calculate $\widehat{S_{yx}(f)}$ are the instantaneous phases at the detector output, sampled and digitized.  This is the case of all the spectra (Figs.~\ref{fig:Sphi-W501}, \ref{fig:Sphi-W501-E5052B}, \ref{fig:Sphi-Citrine} and \ref{fig:FSWP-Re-Im}) and the coefficient $\mathsf{b}_0$ (Fig.~\ref{fig:b0-W501} and \ref{fig:b0-Citrine}).

\section{Experiments and Results}
\graphicspath{{./Figs/}}

The experiment consists of the measurement of the white noise floor after inserting various dissipative attenuators in the path from the oscillator under test to the phase noise analyzer, as shown on Figure~\ref{fig:Experiment}.

\begin{figure*}[t]
{\Large\sffamily (A): all spectra overlapped}\\[-1.5em]
\begin{center}\includegraphics[scale=0.38]{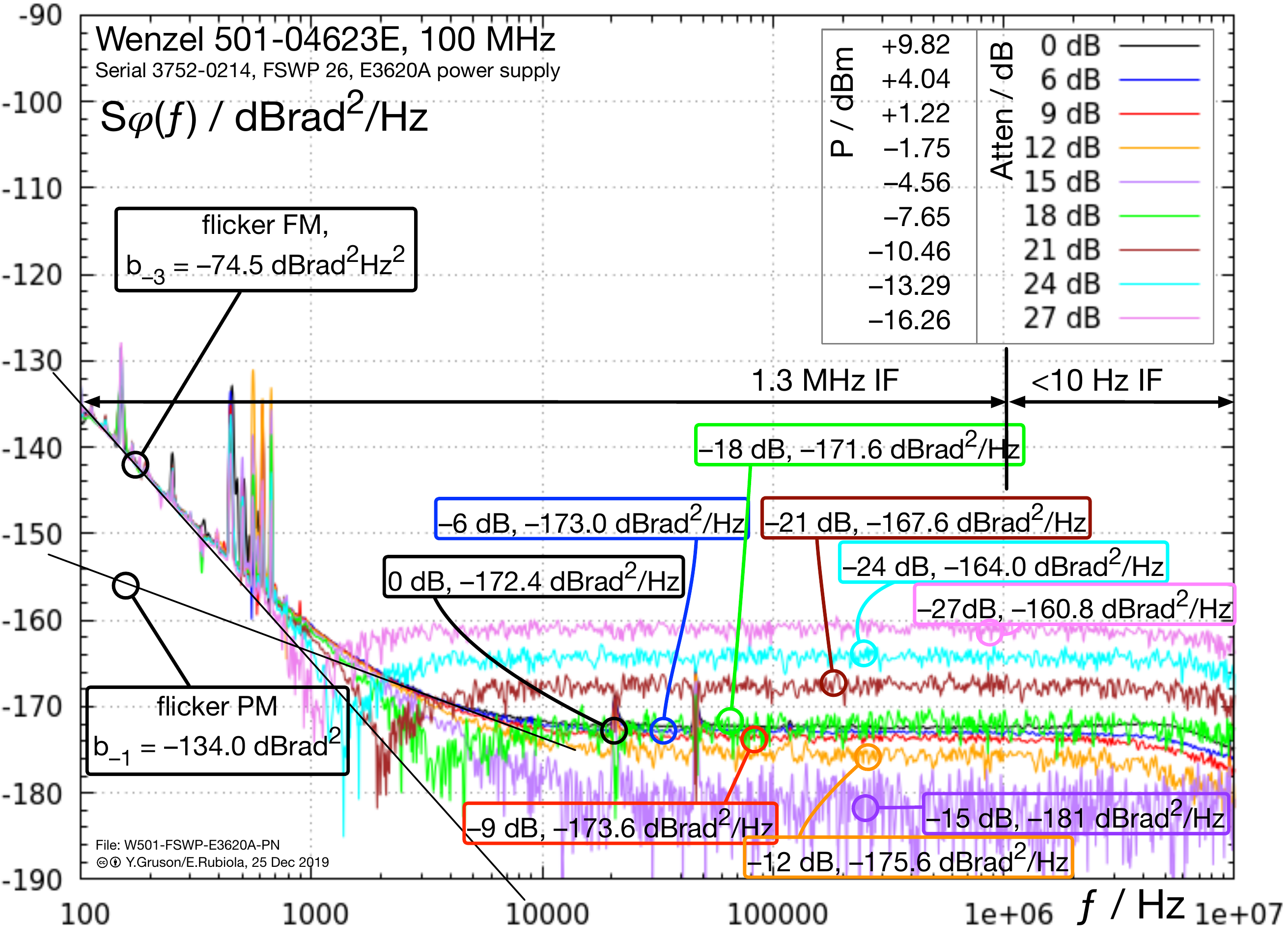}\end{center}\par~\par~\par
{\Large\sffamily (B): Same spectra of (A), fitted with the model (see Section~\ref{sec:Interpretation}).}\\[-1.5em]
\begin{center}\includegraphics[width=0.99\textwidth]{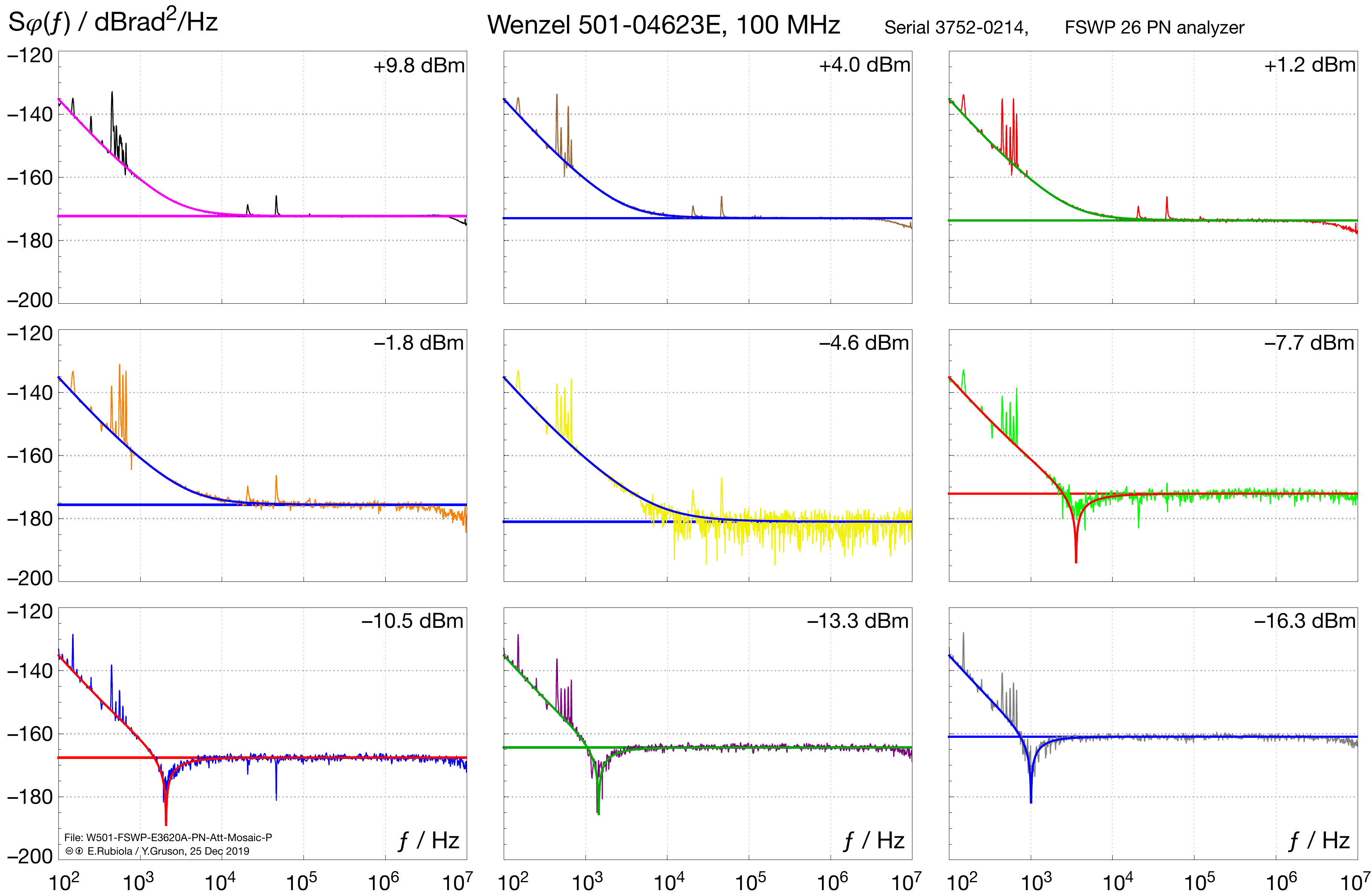}\end{center}\par
\caption{Phase noise of the Wenzel 501-04623E OCXO measured with the FSWP 26.}
\label{fig:Sphi-W501}
\end{figure*}


\begin{figure}[t]
\centering
\includegraphics[width=3.5in]{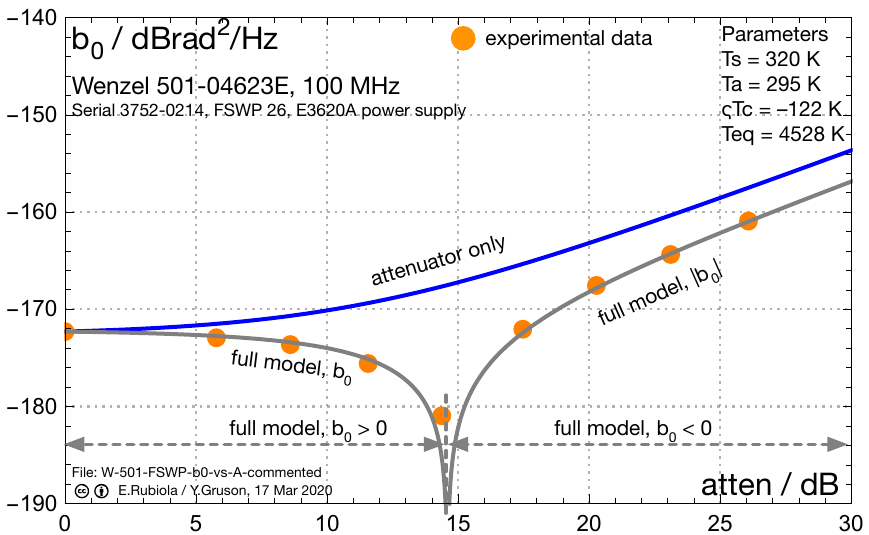}
\caption{White noise floor $\mathsf{b}_0$ (dots) taken from Fig.~\ref{fig:Sphi-W501}-A, compared to the ``attenuator only'' model based on \req{eqn:b0-Atten}.  The ``full model'' plot and the parameters are discussed in Section~\ref{sec:Interpretation}.}
\label{fig:b0-W501}
\end{figure}

Two oscillators are tested, $\mathcal{A}$ a Wenzel 501-04623E, and $\mathcal{B}$ a Wenzel 501-25900B ``Golden Citrine,'' both\ 100-MHz OCXOs intended for the lowest-noise applications.  The former dates more than 20 years ago.  The latter is the top low-PM-noise oscillator by Wenzel.  After comparing to the spectra published on the web pages of several manufactures, $\mathcal{B}$ is the OCXO that exhibits the lowest white noise we have found, below  $-190$ dBc/Hz \cite{Wenzel-Citrine}.  

The oscillator is clamped on a vibration-damping breadboard, of the same type commonly seen in optical experiments.  A 150-MHz low-pass filter (MiniCircuits SLP-150) is inserted at the oscillator output. The attenuation is obtained by stacking small-size SMA attenuators at the filter output, close to the oscillator.  The attenuators (Radiall brand) are intended for DC to 18 GHz.  In most of the tests, the phase-noise analyzer is a Rohde Schwarz FSWP 26 with high-stability OCXO and cross-spectrum options.  
The phase-noise analyzers are referenced to a T4Science Hydrogen maser, in turn monitored vs other masers of the same type.  The power is measured with a Rohde \& Schwarz power meter, which replaces temporarily the phase-noise analyzer before each measurement.  
The attenuation is evaluated as the power ratio.  
All the experiments are done in a Faraday cage with usual isolation transformer and EMI filters.  Temperature and humidity are stabilized to  $22\pm0.5\,\Celsius$ and  $50\,\% \pm10\,\%$  by a PID control, which also guarantee a drift smaller than 0.2 K/hour.  The environment control is probably overdone for PM noise measurements, yet it helps to get conservative results.

\begin{figure}[t]
\centering
\includegraphics[width=3.5in]{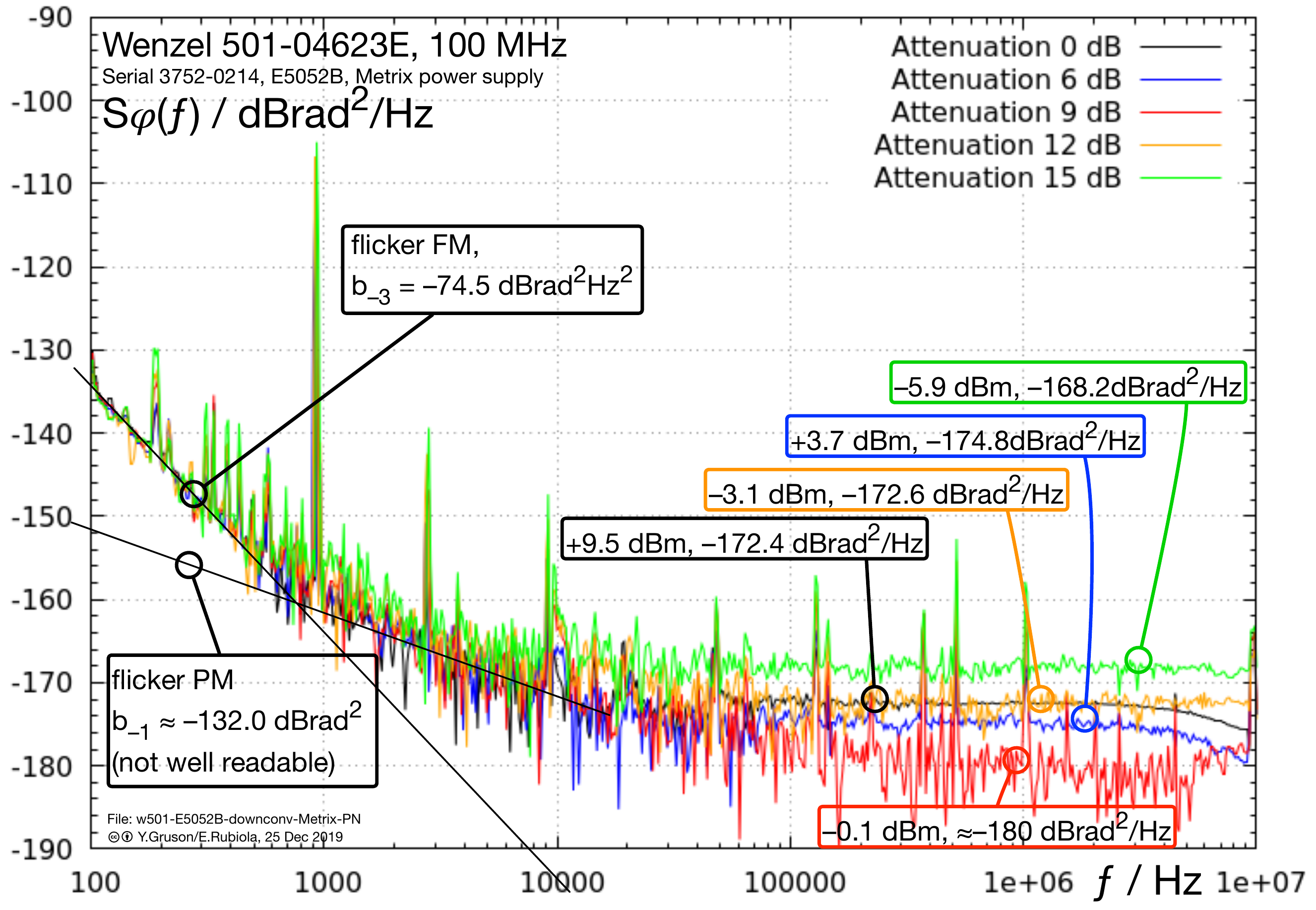}
\caption{Phase noise spectrum of the same oscillator of Figure \ref{fig:Sphi-W501}, measured with a Keysight E5052B phase noise analyzer.}
\label{fig:Sphi-W501-E5052B}
\end{figure}

\begin{figure*}[t]
{\Large\sffamily (A): all spectra overlapped}\\[-1.5em]
\begin{center}\includegraphics[scale=0.4]{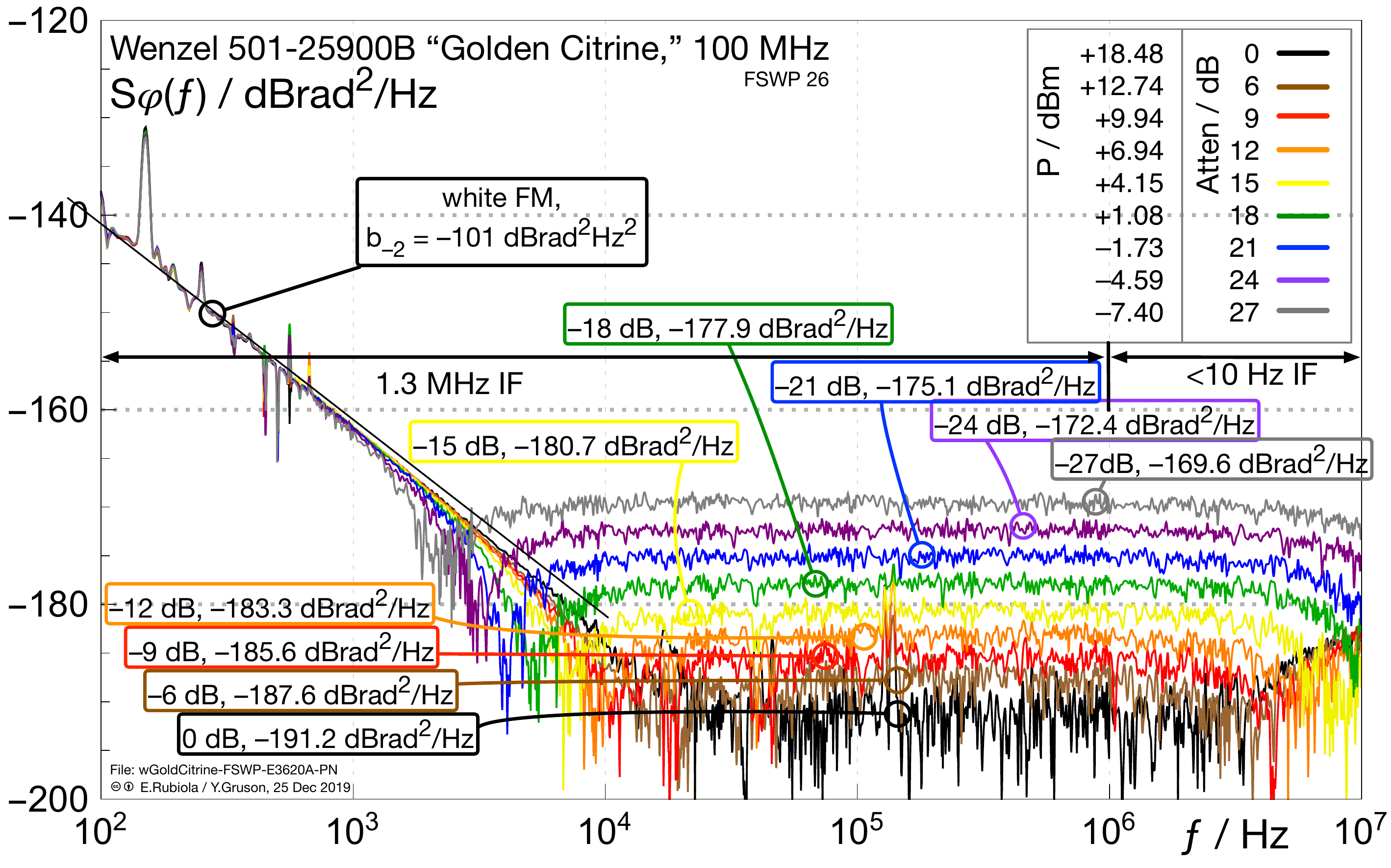}\end{center}\par~\par
{\Large\sffamily (B): Same spectra of (A), fitted with the model (see Section~\ref{sec:Interpretation}).}\\[-1.5em]
\begin{center}\includegraphics[width=0.99\textwidth]{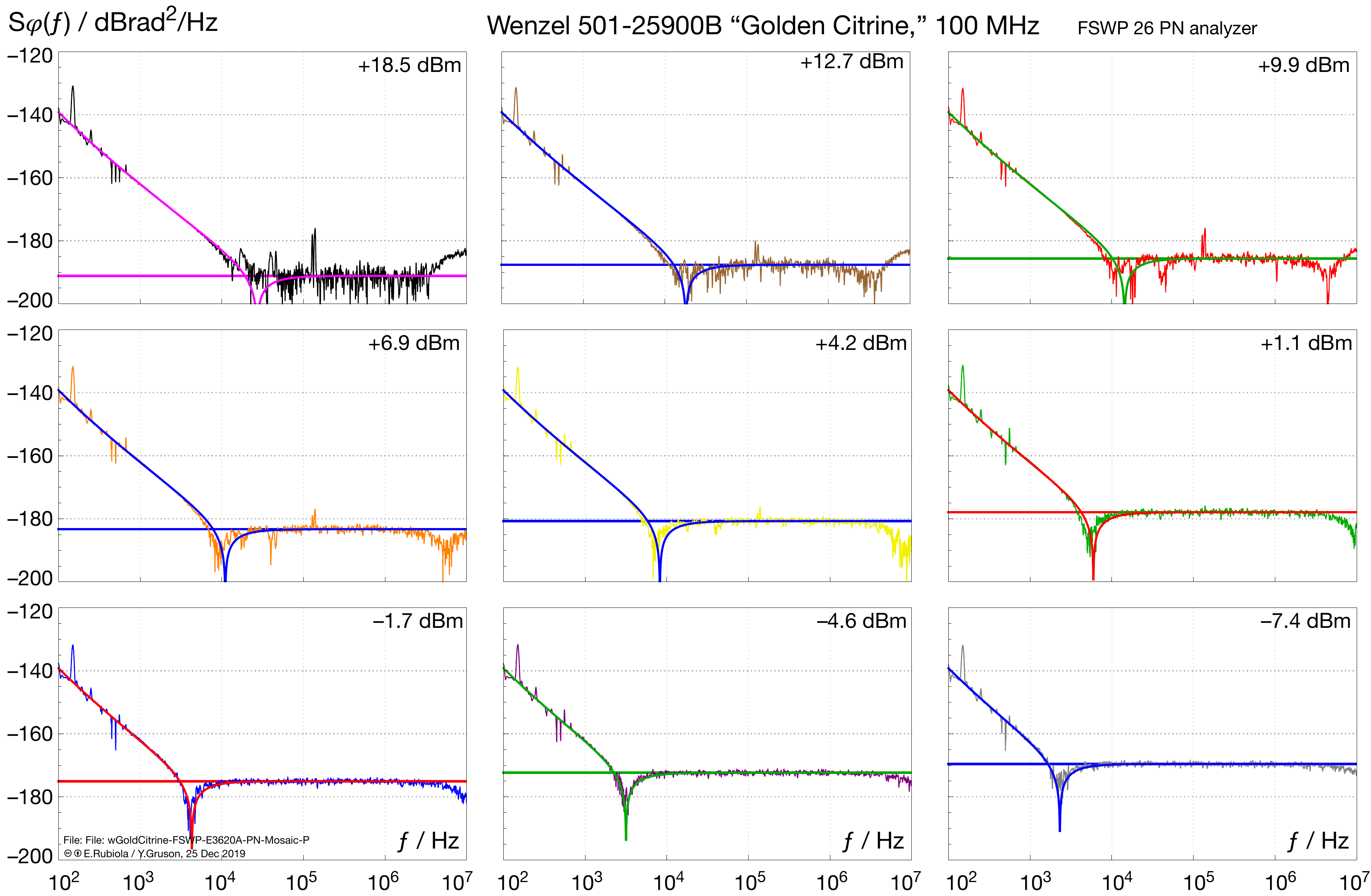}\end{center}\par
\caption{Phase noise  of a Wenzel 501-25900B ``Golden Citrine'' 100 MHz OCXO measured in the same conditions and with the same instruments of Fig.~\ref{fig:Sphi-W501}.}
\label{fig:Sphi-Citrine}
\end{figure*}

Figure~\ref{fig:Sphi-W501} shows the phase noise spectra of the oscillator $\mathcal{A}$, observed\ with different values of the attenuation between 0 dB and 27 dB\@.  
The experimental data (dots) on Fig.~\ref{fig:b0-W501} are the white PM noise from Fig.~\ref{fig:Sphi-W501}, averaged on a suitable region 2--3 decades wide.  
Surprisingly, the observed floor does not match the ``attenuator only'' plot.  The latter is calculated from \req{eqn:b0-Atten}.  Instead, the floor decreases monotonically from 0 dB to 15 dB attenuation, and it increases monotonically beyond.  

Measuring the oscillator $\mathcal{A}$ with a Keysight E5052B, we see that the white PM noise decreases monotonically with the attenuation, attends a minimum at 9 dB, and increases at higher attenuation (Fig.~\ref{fig:Sphi-W501-E5052B}). We could not push the attenuation beyond 15 dB because the carrier power falls below the minimum for the E5052B\@.

The anomalously low white PM noise when an attenuator is introduced was first observed by one of us (AR) in his radio amateur lab at home, measuring a Wenzel 501-04538F 10 MHz OCXO with a FSWP 8.

Comparing Fig.~\ref{fig:Sphi-W501-E5052B} to Fig.~\ref{fig:Sphi-W501}-A, 
the calibration of the two instruments is consistent within at most a small fraction of a dB\@.  The flicker of frequency is the same, $\mathsf{b}_{-3}=-74.5$ \unit{dB\,rad^2Hz^2}.  Likewise, the white noise floor at 0 dB attenuation is the same, $\mathsf{b}_0=-172.4$ \unit{dB\,rad^2/Hz}.  
The 2-dB discrepancy in the flicker PM noise is not significant because the $\mathsf{b}_{-1}$ coefficient is hardly readable on Fig.~\ref{fig:Sphi-W501-E5052B}.

\begin{figure}[t]
\centering
\includegraphics[width=3.5in]{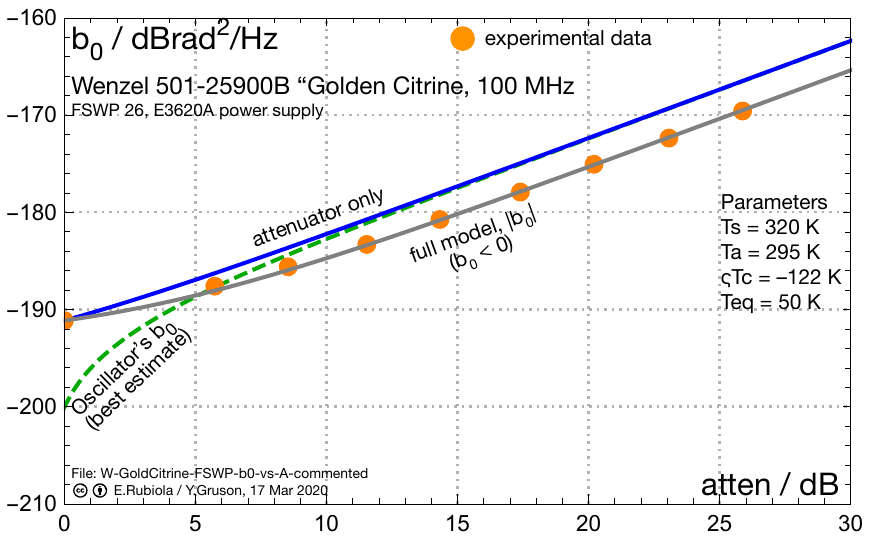}
\caption{White noise floor $\mathsf{b}_0$ (dots) taken from Fig.~\ref{fig:Sphi-Citrine}-A, compared to the ``attenuator only'' model based on \req{eqn:b0-Atten}.  The ``full model'' plot and the parameters are discussed in Section~\ref{sec:Interpretation}.}
\label{fig:b0-Citrine}
\end{figure}

Figure~\ref{fig:Sphi-W501}-B shows the same plots of Fig.~\ref{fig:Sphi-W501}-A, just separated for better readability.  The most interesting fact is the appearance of dips at 1--1.5 kHz for attenuation of  $\geq18 $ dB\@.   

Figures~\ref{fig:Sphi-Citrine} and \ref{fig:b0-Citrine} refer to the same experiment of Fig.~\ref{fig:Sphi-W501} and \ref{fig:b0-W501}, but for the oscillator $\mathcal{B}$.  In this case the white noise floor increases monotonically with the attenuation, but there is a significant discrepancy between the experimental data and the ``attenuator only'' floor predicted by \req{eqn:b0-Atten}.  Additionally, dips are seen on Fig.~\ref{fig:Sphi-Citrine}-B at 2--20 kHz, more noticeable than on Fig.~\ref{fig:Sphi-W501}-B\@.

\begin{figure}[t]
\centering
\includegraphics[width=3.5in]{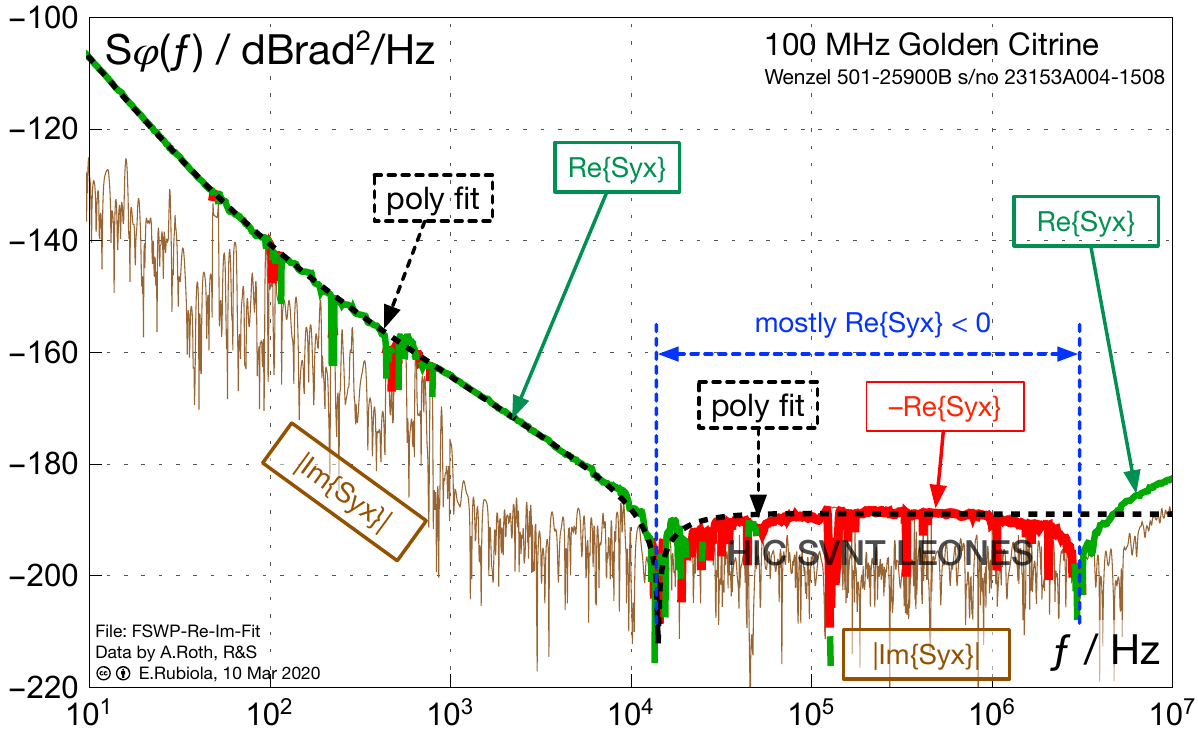}
\caption{PM noise of a ``Golden Citrine'' 100-MHz oscillator, measured with a hacked FSWP\@.  
The polynomial fit gives $\mathsf{b}_{-4}=1.51{\times}10^{-7}$ \unit{rad^2\,Hz^3} ($-68.2$ dB),
$\mathsf{b}_{-3}=3.8{\times}10^{-9}$ \unit{rad^2\,Hz^2} ($-84.2$ dB),
$\mathsf{b}_{-2}=3.39{\times}10^{-11}$ \unit{rad^2\,Hz} ($-104.7$ dB),
$\mathsf{b}_{-1}=-6.31{\times}10^{-16}$ \unit{rad^2} ($|\mathsf{b}_1|=-152$ \unit{dB\,rad^2}), and $\mathsf{b}_{0}=-1.26{\times}10^{-19}$ \unit{rad^2/Hz} ($|\mathsf{b}_{0}|=-189$ \unit{dB\,rad^2/Hz}).
The region where $\Re\{S_{yx}\}<0$ is not a valid PM-noise spectrum, but $-\Re\{S_{yx}\}$  provides useful information related to $\left|S_{yx}\right|$ because $\Re^2\{S_{yx}\}\gg\Im^2\{S_{yx}\}$ almost everywhere.  The ancient-Latin expression HIC SVNT LEONES, usually translated as ``here be dragons,''  refers to an unexplored land, or to a land where humans are not permitted.}
\label{fig:FSWP-Re-Im}
\end{figure}

Inspired by the theory (Sec.~\ref{sec:Cross-spectrum}), we hacked a FSWP at the Rohde Schwarz R\&D facility in München, extracting $\Re\{\left<S_{yx}(f)\right>\}$ and $\Im\{\left<S_{yx}(f)\right>\}$.  This instrument is of the same type of that we have in Besancon.  In München we measured a third oscillator $\mathcal{C}$, a 100-MHz Wenzel 501-25900B ``Golden Citrine'' OCXO, same brand and type of $\mathcal{B}$.  The DUT is connected via a 3-dB attenuator, and the FSWP had internal 5-dB attenuation mechanically switched for better impedance matching.  Additionally, there is a 2.4 dB (typical) loss inside the FSWP, before the power splitter.  All losses accounted for, the signal level at the power splitter input is 8.4 dBm, measured with the internal power meter.  
The result is shown on Fig.~\ref{fig:FSWP-Re-Im}.  The phase noise is represented as $\Re\{\left<S_{yx}\right>\}$.  The negative, invalid outcomes are replaced with $-\Re\{\left<S_{yx}\right>\}$ and shown in different color.  The quantity $|\Im\{\left<S_{yx}\right>\}|$  gives an indication about the averaging limit of the instrument.  Because $|\Im\{\left<S_{yx}\right>\}|\ll|\Re\{\left<S_{yx}\right>\}|$ almost everywhere in the spectrum, $|\Re\{\left<S_{yx}\right>\}|$  is a good approximation of  $|\left<S_{yx}\right>|$.

\section{Interpretation}\label{sec:Interpretation}
The dips found at 1--1.5 kHz in Fig.~\ref{fig:Sphi-W501}, and also at 2--20 kHz in Fig.~\ref{fig:Sphi-Citrine}, suggest that $S_\varphi(f)$ changes sign at these points, being $S_{\varphi}(f)>0$ for $f<f_\text{dip}$, and $S_{\varphi}(f)<0$ beyond.  The sign change occurs because of $\varsigma S_d$ in \req{eqn:SyxD}, related to the fact that the displayed $S_\varphi(f)$ is actually $|\left<S_{yx}(f)\right>|$, where $x$ and $y$ are the phase of the DUT measured by the two channels inside the instrument.  The absolute value turns the sign-change into the sharp dip observed on the log scale.
This is experimentally confirmed in Fig.~\ref{fig:FSWP-Re-Im}. 
By the way, the presence of such dips was already predicted by a simulation in \cite[Fig.~1(b) and Fig.~3]{Nelson-2014}.

From a theoretical standpoint, the combined effect of the attenuator \req{eqn:b0-Atten} and of the power splitter \req{eqn:bo-Splitter} results in 
\begin{equation}
\mathbb{E}\left\{\mathsf{b}_{o}\right\} = 
\frac{kT_i}{P_i} + \frac{k(1-A^2)T_a}{A^2P_i} - \frac{kT_s}{A^2P_i}
\label{eqn:boE}
\end{equation}
at the attenuator output.  This contains two systematic effects: the attenuator noise (positive), and the thermal energy of the power splitter (negative).  At high attenuation ($A^2\rightarrow0$), the RF spectrum associated to the noise sidebands tends to $kT_{a}$.  In this condition, \req{eqn:boE} predicts $\mathrm{b}_{o}<0$ because the temperature $T_{s}$ inside the instrument is obviously higher than the attenuator (and room) temperature $T_{a}$. 

Let us start from the oscillator $\mathcal{A}$, the old Wenzel 501-04623E (Fig.\ref{fig:b0-W501}).   Using the absolute-value estimator, the expected $\mathsf{b}_{o}$ is 
\begin{align}
\mathbb{E}\{\widehat{\mathsf{b}_{o}}\} &= 
\left|\frac{kT_i}{P_i}+\frac{k(1-A^2)T_a}{A^2P_i}-\frac{kT_s}{A^2P_i}\right|~.
\label{eqn:bo-Estim}
\end{align}
Fitting the experimental points with \req{eqn:bo-Estim} fails because there results a too high $T_{s}$.  
Because the isolation between channels cannot be perfect, we replace $T_{s}$ with $T_{s}-\varsigma T_{c}$, where $\varsigma T_{c}$ expresses the crosstalk given in terms of a temperature, and $\varsigma$ has the same meaning as in \req{eqn:SyxD}.  Accordingly, \req{eqn:bo-Estim} rewrites as
\begin{align}
\mathbb{E}\{\widehat{\mathsf{b}_{o}}\} &= 
\left|\frac{kT_i}{P_i}+\frac{k(1-A^2)T_a}{A^2P_i}+\frac{k(\varsigma T_c-T_s)}{A^2P_i}\right|~.
\label{eqn:bo-Abs-Xtalk}
\end{align}
Notice that there are two unknowns in \req{eqn:bo-Abs-Xtalk}, $T_i$ and $\varsigma T_c-T_s$.  The former is dominant at no attenuation ($A^2=1$), where the observed PM noise is rather high.  The latter is dominant at high attenuation ($A^2\rightarrow0$).  Because $\varsigma T_c-T_s$ appears as a single quantity in \req{eqn:bo-Abs-Xtalk}, separating $\varsigma T_{c}$ from $T_{s}$ is somewhat artificial, but it is useful in that it provides physical insight.  We assume $T_{a}=295$ K (23 \Celsius) and $T_{s}=320$ K (47 \Celsius) a convenient round number quite plausible for the instrument inside.  Fitting the data of Fig.~\ref{fig:b0-W501} with \req{eqn:bo-Abs-Xtalk} results in $T_{i}=4528$ K and $T_{c}=122$ K\@.  This is the curve labeled ``full model.''  Using $\mathsf{b}_{i}=kT_{i}/P_{i}$, with $P_{i}=9.6$ mW ($+9.8$ dBm at $A^2=1$), we get $\mathsf{b}_{i}=6.5{\times}10^{-10}$ \unit{rad^2/Hz}  ($-171.9$ \unit{dB\,rad^2/Hz}).   Comparing this value to the readout ($-172.4$ \unit{dB\,rad^2/Hz} at $A^2=1$), the instrument introduces a bias of $-0.5$ dB due to the combined effect of power splitter and crosstalk.

Removing the absolute value in \req{eqn:bo-Abs-Xtalk} yields
\begin{align}
\mathbb{E}\{\widehat{\mathsf{b}_{o}}\} & = 
\frac{kT_i}{P_i}+\frac{k(1-A^2)T_a}{A^2P_i}+\frac{k(\varsigma T_c-T_s)}{A^2P_i}~,
\label{eqn:bo-Xtalk}
\end{align}
which results in $\mathsf{b}_{o}>0$ up to 15 dB attenuation, and in $\mathsf{b}_{o}<0$ beyond.

Rewriting the polynomial model \req{eqn:plaw} for the absolute-value estimator we get
\begin{equation}
\mathbb{E}\{\widehat{S_{\varphi}(f)}\} = \left|
\frac{\mathsf{b}_{-3}}{f^3} +
\frac{\mathsf{b}_{-2}}{f^2} +
\frac{\mathsf{b}_{-1}}{f} +
\frac{kT_i}{P_i}+\frac{k(1-A^2)T_a}{A^2P_i}+\frac{k(\varsigma T_c-T_s)}{A^2P_i}
\right|\,.
\label{eqn:Syx-Abs-Xtalk}
\end{equation}
Evaluating \req{eqn:Syx-Abs-Xtalk} with $\mathsf{b}_{-3}=3.5{\times}10^{-8}$ \unit{rad^2Hz^2} ($-74.5$ \unit{dB\,rad^2Hz^2}), 
$\mathsf{b}_{-2}=0$, and $\mathrm{b}_{-1}=4{\times}10^{-14}$ \unit{dB\,rad^2} ($-134$ \unit{dB\,rad^2}), taken from Fig.~\ref{fig:Sphi-W501}-A, we find the solid lines overlapped to the experimental spectra of Fig.~\ref{fig:Sphi-W501}-B\@.  The model matches the experiment, and predicts precisely the dips.  These dips occurr at $\geq18$ dB attenuation, where $\mathsf{b}_{o}<0$.

Now we turn our attention to the oscillator $\mathcal{B}$, the Wenzel 501-25900B ``Golden Citrine.''.  Looking at Fig.~\ref{fig:Sphi-Citrine}-A and Fig.~\ref{fig:b0-Citrine}, we notice that the white noise floor increases monotonically increasing the attenuation, and the dips are present for all the values of the attenuation --- albeit these dips are not clear at 0 dB and 6 dB because of insufficient averaging.  This indicates that $\mathsf{b}_{o}<0$ in all cases.  Evaluating \req{eqn:Syx-Abs-Xtalk} with the same $T_{a}$, $T_{s}$ and $T_{c}$ as above, we find $T_\text{eq}=T_{i}=50$~K\@.  The model fits well the experimental data, as shown on Fig.~\ref{fig:b0-Citrine}.  The temperature of 50 K is equivalent to a white noise floor of $-200$ \unit{dB\,rad^2/Hz} at $+18.5$ dBm (70.5 mW) output power, with no attenuation.  Finally, \req{eqn:Syx-Abs-Xtalk} predicts precisely the dips seen on at Fig.~\req{fig:Sphi-Citrine}-B\@.

\subsection{The Origin of the Crosstalk}\label{sec:Crosstalk}
Trying to understand the crosstalk, we look at the part of the FSWP where the strongest and the weakest signals come close to one another, which is the input mixer.
Let us put numbers together with this idea.  For linear conversion, the LO signal should not be lower than $+20$ dBm.  The phase noise of a state-of the art synthesizer at 100 MHz carrier is of the order of $-160$  \unit{dB\,rad^2/Hz}.  For reference, the R\&S SMA100A synthesizer with the low-phase-noise option SMA-B22 has a white floor of this order \cite[data sheet, p.~12]{RS:SMA100A}.
At $+20$ dBm power, the white-noise sidebands are of $-140$ dBm/Hz, that is, $10^{-17}$ W/Hz.  The crosstalk $kT_{c}$ we search for is of $1.7{\times}10^{-21}$ W/Hz with $T_{c}=122$ K\@.  This is 38 dB smaller than the LO sidebands.  A coupling of the order of $-38$ dB due to leakage is quite plausible for a good mixer circuit.  
Besides, the absence of discontinuity in the spectrum (Fig.~\ref{fig:Sphi-W501} and \ref{fig:Sphi-Citrine}) at 1 MHz indicates that the crosstalk does not depend on the operating mode, which excludes some other parts of the instrument.  The presence of a small amount of anticorrelated flicker PM is also possible, for the same reason.
Anyway, this interpretation is just a guess, not based on the internal design nor on specific measurements.

\subsection{Inside the Oscillator}\label{sec:Oscillator}
\begin{figure}[t]
\centering
\includegraphics[width=3.5in]{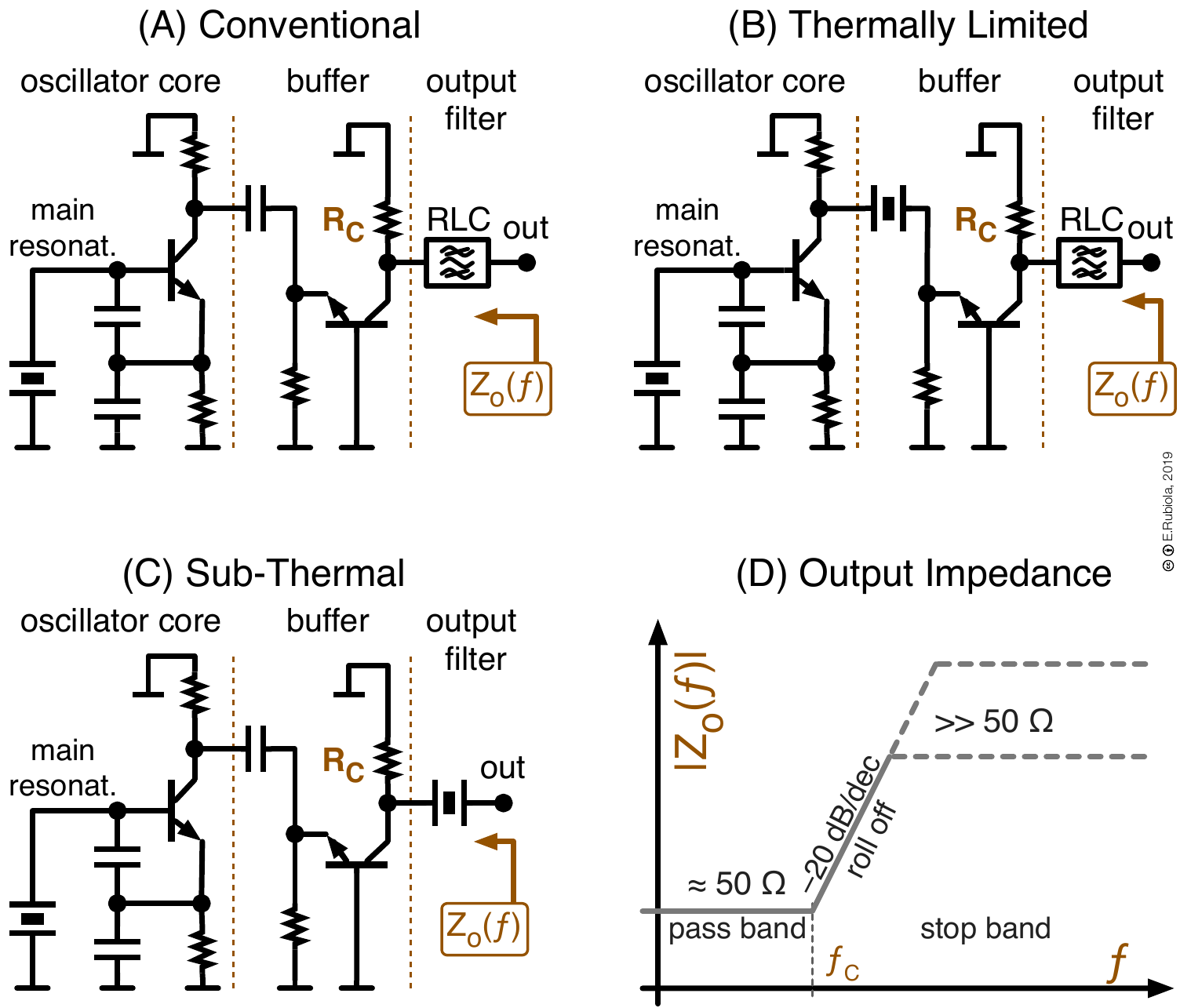}
\caption{Simplified scheme of the low-noise quartz oscillators.  The key point is the interplay of filters and impedances.  Otherwise, commercial oscillators may differ from the schemes shown.}
\label{fig:Oscillators}
\end{figure}

We address the question of the origin of $T_\text{eq}$, and why it can be smaller than the room temperature.  From our purposes, the oscillator consists of a core (the auto-oscillator in strict sense), a buffer, and an output filter (Fig.~\ref{fig:Oscillators}).  The attenuation in the filter stopband is generally achieved by reflecting the power back to the generator's internal impedance.

The \emph{\bfseries conventional} oscillators may have a lowpass or bandpass RLC filter at the output to suppress the harmonic distortion and to solve other practical problems (Fig.~\ref{fig:Oscillators}-A).  Such filter cannot have a bandwidth smaller than a few MHz at 100 MHz carrier because the quality factor $Q$ of these resonators is of the order of 10--20 in practical conditions.  As a consequence, the output impedance is reasonably matched in the whole Fourier-frequency span, and the white noise is chiefly the noise of the sustaining amplifier, where the carrier is the weakest.

In the \emph{\bfseries thermally limited} quartz oscillator, a quartz resonator is present between the core and the buffer.  Such filter can be the main resonator if the carrier is extracted from the resonator's ground pin \cite{Rohde-1975}, \cite[Fig.~4-58 to 4-61]{Rohde-1997}, or a second quartz resonator (Fig.~\ref{fig:Oscillators}-B).  Out of the resonator bandwidth $\nu_0(1\pm1/2Q)$, the quartz is a high impedance circuit, thus the noise of the sustaining amplifier is not transmitted to the buffer.  
The noise associated to the resonator's motional resistance is also rejected, for the same reason. 
The buffer (a common-base amplifier) has low input impedance and low noise figure, thus the white noise is chiefly limited by the physical temperature of the collector resistor $R_C$ at the output.  Such oscillators may have an additional RLC output filter of the same type discussed before.

In the \emph{\bfseries sub-thermally limited} quartz oscillator, a quartz resonator or a quartz filter is introduced in series to the output \cite[Fig.~7]{Rohde-1978}, with no further amplification.  The output filter has a small cutoff frequency even with the low $Q$  imposed by the heavy load condition.  For example, taking  $Q=5000$ at 100 MHz, the cutoff frequency is $f_{c}=10$ kHz.  
For comparison, a good resonator at this frequency has $Q>10^{5}$, unloaded.
Out of the bandwidth $\nu_0(1\pm1/2Q)$, the output impedance is quite high ($|Z_{o}|\gg50~ \Omega$), which gives the appearance of a \emph{cold}\ source.  There no violation of the second principle because the filter is obviously in thermal equilibrium with the environment.  However, the \emph{electrical access} to the thermal energy is open.  In this condition, the input power splitter of the noise analyzer is reasonably well matched only in the pass band, and nearly open circuit in the stopband.  In the stopband, the expected cross spectrum relates to the thermal energy of the power splitter (and to the crosstalk, if any), wich has negative sign in the correlation.  

Simple attempts to measure the output impedance failed because the impedance analyzers do not work in the presence of the strong carrier at the input.  
We disassembled two 100 MHz oscillators, a Wenzel 501-04623E and a Wenzel Citrine, the same type as the oscillator $\mathcal{A}$ and $\mathcal{B}$, respectively.  The oscillator $\mathcal{A}$ is of the conventional type, with a RLC filter at the output.  The white PM noise limited by the signal-to-noise ratio in the sustaining amplifier.  The oscillator $\mathcal{B}$ is of the sub-thermally limited type, with a quartz resonator in series to the output.  Albeit we did not reverse-engineer the oscillator, the two values of $T_\text{eq}$, 4528 K and 50 K, are consistent with the oscillator architecture.

\section{Discussion}\label{sec:Discussion}
We have seen that the (anti-)correlated noise inside the instrument can be modeled as a temperature, which is $\varsigma T_c-T_s$.  Let us look at $T_{s}$ and $T_{c}$ separately.
Because \req{eqn:bo-Splitter} is based on simple and well-established physics, a software correction inside the instrument can compensate for $T_{s}$ in a reliable way.  An accuracy of a few kelvins is all what is needed. Likewise, \req{eqn:b0-Atten} can be used to correct for the effect of a switchable dissipative attenuator, if present.
By contrast, there is no general way to compensate for $T_{c}$.
We have no a priori reason to trust it as a constant in the carrier-frequency range (4 decades), nor as reproducible parameter across different specimens or architectures.
The brute force approach of putting the power splitter in a liquid-He cryostat \cite{Hati-2017} is not effective because of the crosstalk.  In our experiment 70\% of the bias error is due to the power splitter, 30\% to the crosstalk.  However, compensating for $T_{s}$ alone is a general solution, and mitigates the problem.  

The real-part estimator $\Re\left\{\left<S_{yx}\right>_{m}\right\}$ is superior to the traditional estimator $\left|\left<S_{yx}\right>_{m}\right|$ in that
(i) it converges faster because the background noise in $\Im\left\{\left<S_{yx}\right>_{m}\right\}$ is not taken in, and 
(ii) it reveals the negative, nonsensical outcomes.

Measuring the oscillator $\mathcal{A}$, the experimentalist may be satisfied of the spectra taken with no attenuation ($A^2=1$) because
\vspace*{-1ex}
\begin{itemize}\addtolength{\itemsep}{-1ex}
\item Two instruments from the major brands, with similar correlation algorithm but radically different in the RF architecture and in the detection principle, are in perfect agreement.
\item The systematic error in the white noise, revealed by our rather complex experiment, is of a mere $-0.5$ dB, not alarming. 
\end{itemize}
\vspace*{-1ex}
Conversely, the white noise floor measured on the oscillator $\mathcal{B}$ is a complete nonsense because $\Re\left\{\left<S_{yx}(f)\right>_m\right\} <0$ inside the instrument.

Unlike most domains of metrology (mass, length, etc.), a PM noise spectrum consists of hundreds or thousands of points on the $S_\varphi(f)$ plot.  The common ditto \emph{too much information is no information} rises the question of the nature of the measurand.  General experience indicates that the polynomial law \req{eqn:plaw} describes well the PM noise spectrum of quartz ad dielectric oscillators, thus a small number (4--5) of  parameters $\mathsf{b}_{n}$ tell the whole story.  In optics, some additional terms appear, like bumps and blue noise, which call for a small number of additional coefficients \cite{Calosso-2016}. Regardless of the model we choose, a small number of missing points, like the negative spurs we have seen, is not a real nuisance and can be ignored.  The experienced scientist does this after visual inspection.  By contrast, an irregular behavior over a wide frequency range has to be taken seriously.

Ultimately, some concepts found in the International Vocabulary of Metrology (VIM) \cite{VIM} and in the Guide to the Expression of Uncertainty in Measurement (GUM) \cite{GUM} (see also \cite{Hall-2016,Possolo-2017}) should be introduced in phase noise measurements.  Going through the VIM, the following definitions are relevant to our experiments: \emph{Type~A} and \emph{Type~B} evaluation of uncertainty (2.28 and 2.29), the \emph{influence quantities} (2.52), the \emph{definitional uncertainty} (2.27), and the \emph{null measurement uncertainty} (4.29). 
Because none of us is a true expert of uncertainty in metrology, subtleties may escape from our attention.
However, this article shows that the assessment of uncertainty in PM noise is still at a too rudimentary stage.
The \emph{following digression} is intended to \emph{stimulate a discussion}, with no intention of stating rules.

The type A uncertainty $u_A$  can be processed by a statistical analysis of the time series, or in our case of a series of spectra.  Conversely, the type B uncertainty $u_B$ can be determined by other means, chiefly the analysis of the system.  The combined uncertainty (GUM Sec.~2.3.4 and Sec.~5) is $u_C = \sqrt{u_{A}^{2}+u_{B}^{2}}$.
In engineering, implicit reference is often made to the \emph{expanded uncertainty} (VIM 2.35), with a coverage probability of 95\,\%. 

Most of our practical knowledge about uncertainty comes from Fig.~\ref{fig:Sphi-W501} to \ref{fig:b0-Citrine}.  
By contrast, Fig.~\ref{fig:FSWP-Re-Im} provides the strong evidence of negative outcomes in a large portion of spectrum.  In turn, Fig.~\ref{fig:FSWP-Re-Im} supports our conclusion that some portions of spectra are correctly interpreted as negative outcomes ($\mathsf{b}_0<0$ on Fig.~\ref{fig:Sphi-W501} to \ref{fig:b0-Citrine}), made positive by the absolute-value estimator.

The single-channel background of the instrument is chiefly zero-average Gaussian noise with white or colored spectral distribution, thus it falls in the type A uncertainty.  Averaging on $m$ spectra, the single-channel is reduced by a factor of $1/\sqrt{n}$ for the absolute value estimator, and $1/\sqrt{2m}$ for the real-part estimator.  The practical impact on the estimation of the white noise can be easily made negligible by averaging on a large $m$, say $10^6$, ultimately limited by the time-frequency uncertainty theorem.

Two contributions to the type B uncertainty are obviously identified, a calibration factor and the correlated effects discussed.  This suggests a minimalist model like $u_B(S_\varphi) = a_1 S_\varphi + a_0$.
The calibration factor $a_1$ is dominant on the left-hand side of the spectrum, where $S_\varphi(f)$ is higher.  Based on the $1/f^3$ noise seen on Fig.~\ref{fig:Sphi-W501} and \ref{fig:Sphi-W501-E5052B} (same oscillator, measured with two instruments of different brand, RF architecture, and principles), we can infer that the uncertainty is not greater than 0.2--0.3 dB, i.e., $\lesssim2\,\%$ in the phase-to-number conversion.  The same is expected on Fig.~\ref{fig:FSWP-Re-Im}.  Anyway such small value is not an issue in the laboratory practice.
By contrast, the noise $k(\varsigma T_c-T_s)/P$ show up in the white noise region, which is our main concern.  
Accepting to go through a complex process, we can measure $k(\varsigma T_c-T_s)/P$ and corrected for it, with a residual uncertainty.  Otherwise, we can use the knowledge gathered to infer the uncertainty in a similar experiment.

\begin{figure}[t]
\centering
\includegraphics[width=3in]{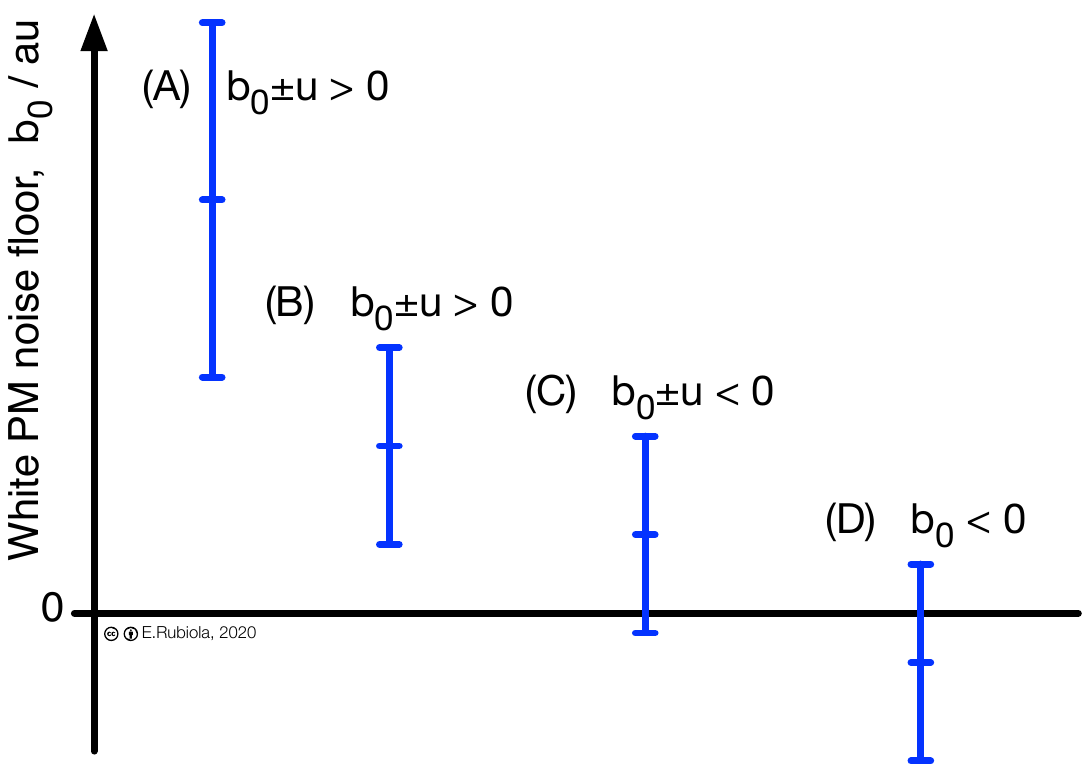}
\caption{Uncertainty concepts, adapted from the VIM \cite{VIM}.}
\label{fig:Zero-uncertainty}
\end{figure}

Here we see the importance of the null measurement uncertainty.  The quantity $k(\varsigma T_c-T_s)/P$, or its uncertainty after correction, sets the detection threshold, i.e., theminimum value of $S_\varphi(f)$ that can be measured.  This concept is illustrated in Fig.~\ref{fig:Zero-uncertainty}.  The cases (A) and (B) comply with rule that $S_\varphi(f) \pm u>0$, with (B) having smaller $u$ as a result of a larger number of averaged spectra (same $u_B$ but smaller $u_{A}$).  By contrast, in (C) the uncertainty bar hits the negative values.  In this case, the correct way to express the result result is $S_\varphi(f)=0$, with a detection threshold $u$.  The case (D) should be discarded at sight, giving the same result $S_\varphi(f)=0$ as in (C).

Let us show the above concepts on the spectrum of Fig.~\ref{fig:FSWP-Re-Im}.  In this case we did only one experiment, thus we have no measure of $\varsigma T_c-T_s$, and no correction for $T_s$ is implemented inside the FSWP\@.  Thus, we take $|\varsigma T_c-T_s|$ rounded to 500 K as a convenient estimate of the instrument limit.   With $P=6.9$ mW ($+8.4$ dBm), $\mathsf{b}_0=kT/P$ gives $u(S_\varphi)=10^{-18}$ \unit{rad^2/Hz}.  Accordingly, the detection threshold is of $-180$ \unit{dB\,rad^2/Hz}.  The forbidden region from 15 kHz to 3 MHz, where $\Re\{S_\varphi(f)\}<0$, corresponds to the case (D) of Fig.~\ref{fig:Zero-uncertainty}. Beyond 3 MHz the spectrum is not trusted, being affected by other limitations.  Anyway, the latter region has little or no practical importance.

Finally, we have seen that the output impedance $Z_{o}(f)$ produces erratic results if it changes significantly in the analysis bandwidth.  This opens the question of whether $Z_{o}(f)$ goes in the definitional uncertainty (it is inside the DUT), it goes in the B-type uncertainty, or if it is an influence quantity.  The role of impedance mismatch is well known in microwave noise measurements \cite{Otoshi-1968, Wedge-1992}, but these concepts have not been transposed to PM noise.

\section{Conclusions}\label{sec:Conclusions}
Our method consists of introducing various values of dissipative attenuation between the oscillator under test and the phase-noise analyzer.  This method is new.  It provides quantitative information about the unwanted correlated effects inside the analyzer, and helps to assess the \emph{null measurement uncertainty}, i.e., the minimum amount of phase noise that can be detected.  
In some circumstances, inserting an attenuator results in lower white PM noise floor.  When this happens, gross errors are around the corner.
The idea that the (anti-)correlated noise inside the instrument can be described in terms of the thermal energy $k(\varsigma T_c-T_s)$ is also new.  This parameter accounts for the temperature of the power splitter at the instrument input, and the crosstalk between the two channels. 

The experiments described provide the evidence that pushing the noise rejection too far by averaging on a large number of data may result in misleading or grossly wrong results.  The reason is in residual correlated effects, not under control.  In general terms, under-estimating the DUT noise is obviously worse than over-estimating it.

Impedance matching in the whole analysis bandwidth is a critical issue.  
Sub-thermally limited oscillators make use of a narrowband reactive filter at the output, which exploits impedance mismatch in the stopband to deliver the lowest noise floor.  However, such filter results in anticorrelated noise due to the thermal energy in the power splitter at the instrument input.
From a different standpoint, the benefit of a sub-thermally limited oscillator is unclear to us if the oscillator is intended to be a part of a system at room temperature.

\section*{Disclaimer}
Our strong statements require an equally strong disclaimer about the commercial products we refer to.  We experimented on them because they were on hand at the right time, as opposite to gathering parts with this research in mind.  By no means we criticize these products, nor we endorse them.  The problems and the inconsistencies we describe relate to unintended, strange, or weird use of these products.  Driven by the genuine scientific curiosity, we share our knowledge with the ultimate intent to contribute to better understanding the physics and the technology of phase noise metrology.  We hope that no misunderstanding will arise, and we  apologize if this will happen.

\section*{Acknowledgments}
This work is funded by the ANR Programme d'Investissement d'Avenir under the Oscillator IMP project (contract ANR-11-EQPX-0033-OSC-IMP) and the FIRST-TF network (contract ANR-10-LABX-48-01), and by the Région Bourgogne Franche Comté.  We thank AR Electronique (the similarity between this brand and the initials of one of the authors is a random outcome), Besancon, for hosting us in their R\&D lab for some additional measurements, and Cristian Bolovan and Radu Ohreac, Rohde \& Schwarz Romania, for generously lending a FSWP 8.

\bibliography{/Users/rubiola/Dropbox/0-Bibliography/Ref-short,/Users/rubiola/Dropbox/0-Bibliography/Ref-Rubiola,/Users/rubiola/Dropbox/0-Bibliography/References,../Ref-local}

\end{document}